\documentclass[10pt, a4paper, twocolumn]{article} 
\usepackage{flushend}

\usepackage[english]{babel} 

\usepackage{microtype} 

\usepackage{amsmath,amsfonts,amsthm} 

\usepackage[svgnames]{xcolor} 

\usepackage[hang, small, labelfont=bf, up, textfont=it]{caption} 

\usepackage{booktabs} 

\usepackage{lastpage} 

\usepackage{graphicx} 

\usepackage{enumitem} 
\setlist{noitemsep} 

\usepackage{sectsty} 
\allsectionsfont{\usefont{OT1}{phv}{b}{n}} 

\usepackage{geometry} 

\usepackage[T1]{fontenc} 
\usepackage[utf8]{inputenc} 

\usepackage{XCharter} 

\usepackage{fancyhdr} 
\fancypagestyle{firstpage}{ 
	\fancyhf{}
}

\newcommand{\authorstyle}[1]{{\large\usefont{OT1}{phv}{b}{n}\color{Black}#1}} 

\newcommand{\institution}[1]{{\footnotesize\usefont{OT1}{phv}{m}{sl}\color{Black}#1}} 

\usepackage{titling} 

\newcommand{\HorRule}{\color{Black}\rule{\linewidth}{1pt}} 

\pretitle{
	\vspace{-30pt} 
	\HorRule\vspace{10pt} 
	\fontsize{32}{36}\usefont{OT1}{phv}{b}{n}\selectfont 
	\color{Black} 
}

\posttitle{\par\vskip 15pt} 

\preauthor{} 

\postauthor{ 
	\vspace{10pt} 
	\par\HorRule 
	\vspace{20pt} 
}

\usepackage{lettrine} 
\usepackage{fix-cm}	

\newcommand{\initial}[1]{ 
	\lettrine[lines=3,findent=4pt,nindent=0pt]{
		\color{Black}
		{#1}
	}{}%
}

\usepackage{xstring} 

\newcommand{\lettrineabstract}[1]{
	\StrLeft{#1}{1}[\firstletter] 
	\initial{\firstletter}\textbf{\StrGobbleLeft{#1}{1}} 
}

\usepackage[backend=bibtex,style=authoryear,natbib=true]{biblatex} 

\usepackage[autostyle=true]{csquotes} 

\title{A Quadruply Lensed SN Ia: Gaining a Time-Delay...Losing a Standard Candle} 

\author{
	\authorstyle{Daniel A. Yahalomi\textsuperscript{1}, Paul L. Schechter \textsuperscript{1,2}, and Joachim Wambsganss\textsuperscript{3}} 
	\newline\newline 
	\textsuperscript{1}\institution{MIT Department of Physics, Cambridge, MA 02139}\\ 
	\textsuperscript{2}\institution{MIT Kavli Institute for Astrophysics and Space Research, Cambridge, MA 02139}\\ 
	\textsuperscript{3}\institution{Zentrum f\"{u}r Astronomie der Universit\"{a}t Heidelberg, Germany} 
}

\date{\today} 

\begin{document}

\maketitle 

\thispagestyle{firstpage} 


\lettrineabstract{We investigate the flux ratio anomalies between macro-model predictions and the observed brightness of the supernova iPTF16geu, as published in a recent paper by \cite{More}. This group suggested that these discrepancies are, qualitatively, likely due to microlensing. We analyze the plausibility of attributing this discrepancy to microlensing, and find that the discrepancy is too large to be due to microlensing alone. This is true whether one assumes knowledge of the luminosity of the supernova or allows the luminosity to be a free parameter. Varying the dark/stellar ratio likewise doesn't help. In addition, other macro-models with quadruplicity from external shear or ellipticity do not significantly improve to model. Finally, microlensing also makes it difficult to accurately determine the standard candle brightness of the supernova, as the likelihood plot for the intrinsic magnitude of the source (for a perfect macro-model) has a full width half maximum of 0.73 magnitudes. As such, the error for the standard candle brightness is quite large. This reduces the utility of the standard candle nature of type Ia supernovae.}


\section{Introduction}
Recently, \cite{Goobar} reported the discovery of a multiply imaged, gravitationally lensed type Ia supernova, iPTF16geu. The supernova was originally detected on September 5, 2016 by the intermediate Palomar Transient Factory (iPTF). Later, observations were made using the Near-Infrared Camera 2 at the Keck telescope (October 22, 2016 and November 5, 2016), in the K$_s$ and J bands, respectively. Optical observations were also made with the Hubble Space Telescope (HST) on October 25, 2016 in three different wavelength filters: F475W, F625W, and F814W. This is an astronomical object of interest because of the standard candle nature of type Ia supernovae. The standard candle feature implies a known intrinsic brightness of the object, and allows us to determine the magnification factor directly, rather than through the lens model. This magnification factor breaks the mass-sheet degeneracy and other degeneracies \citep{Kolatt}. In studying strongly lensed SN Ia, if an accurate macro-model is created, then the time-delay values from the lens model and the intrinsic magnitude and magnification values of the SN Ia can be compared, providing tight constraints on the Hubble parameter.

Subsequently, \cite{More}, presented models for the lensing galaxy. However, there are significant flux ratio anomalies between the observed and predicted flux values of the four lensed images of the supernova. Specifically,  the observation for image A was approximately 2 magnitudes brighter than the macro-prediction, and the observation for image D was approximately 1.5 magnitudes fainter than the macro-prediction. Images B and C had less significant flux ratio anomalies. \cite{More} suggested that these flux ratio anomalies were likely due to microlensing of the foreground galaxy. In this paper, we investigate the likelihood for microlensing to create the deviation in the flux values between the macro-model and the observations for this system.

\section{Background}
\subsection{Gravitational Macrolensing}
Albert Einstein's theory of General Relativity describes gravity as the warping of space-time. It is known that light follows the null trajectory of the geodesic. As gravitational potential warps space-time, it alters the null geodesic, acting as an effective lens for the path that light follows. The effective lensing nature of gravity was observed as early as 1919, when Arthur Eddington observed the deflection of light rays from stars as they passed near the sun. Eddington's observations supported the predictions of General Relativity, refuting those of Newtonian Mechanics. 

Fermat's principle states that the path that light takes between two points is always a stationary point in time. In most situations, light travels on the minimum-time paths, but, light can also travel along maximum-time and saddle-point paths. There are two sources of time delay when light is gravitationally lensed: geometric, due to the extra distance travelled, and gravitational, due to the warping caused by the presence of the lensing mass \citep{BlandfordNarayan}. If we define the angular position of the source as $\theta_S$, the angular position of the lensed image as $\theta_I$, and absorb into a single parameter, $\alpha$, other constants of the setup (such as distances between the observer, lens, and source, and the redshift), then the geometric time delay can be written in the form:

\begin{equation}
	t_{geom}(\theta_I, \theta_S) = \alpha ({\theta_I - \theta_S})^2
	\label{eq:GeomTimeDelay}
\end{equation}

We can then take the three-dimensional relativistic gravitational potential, and integrate it along the line of sight in order to obtain a two-dimensional potential, $\psi(\theta_I)$. We can absorb into a single paramater the other constants of the setup (such as distances between the observer, lens, and source, and the redshift). It has been shown that this single paramater is equal to 2$\alpha$ \citep{BlandfordNarayan}, and the gravitational time delay can be written in the form:

\begin{equation}
	t_{grav}(\theta_I) = 2\alpha (-\psi(\theta_I))
	\label{eq:GravTimeDelay}
\end{equation}

Combining these two contributions to time delay and setting the 2$\alpha$ term equal to unity, we get an equation for total time delay in cosmological units. We use $\tau$ now to note the change in units:

\begin{equation}
\begin{gathered}
	\tau_{total}(\theta_I, \theta_S) = \tau_{geom} + \tau_{grav}\\
	\tau_{total}(\theta_I, \theta_S) = \frac{1}{2}({\theta_I - \theta_S})^2 - \psi(\theta_I)
\label{eq:GravTimeDelay}
\end{gathered}
\end{equation}

Without any gravitational potential, from Fermat's principle, the light will be at the stationary point of the time delay, located at the minimum at $\theta_I = \theta_S$, as expected. Once a gravitational potential is introduced to the system ($\psi < 0$), new extrema can be created if the potential is large enough. Multiple stationary points (maxima, minima, and saddle points) result in a multiply imaged source, an effect called strong lensing. When solved, Fermat's principle predicts an odd number of images, and based on the sizes of observed gravitational potentials astronomically, we expect either three or five images from strong gravitational lensing. The central image cannot be seen, as it is highly demagnified by the lensing galaxy, so we expect to observe either two or four images \citep{SchneiderKochanekWambsganss}.

Gravitational potentials also distort astronomical sources. The images are distorted in both shape and magnification. The shape is distorted by shear $\gamma$, which is the tidal gravitational field or the warping due to surrounding galaxies. The magnification is caused both by the shear and by convergence $\kappa$, which is the local dimensionless matter density. Splitting the shear into real and imaginary parts ($\gamma = \gamma_1 + i\gamma_2$), the distortion of astronomical images is described by the Jacobian Matrix \citep{SchneiderKochanekWambsganss}:

\begin{gather}
 \mathcal{A}(\theta)
 =
  \begin{bmatrix}
   1 - \kappa - \gamma_1 &
   -\gamma \\
   -\gamma &
   1 - \kappa + \gamma_1  
   \end{bmatrix}
\end{gather}

\subsection{Gravitational Microlensing}

Galaxies are composed of a combination of stellar mass and smoothly distributed mass (dark matter). Gravitational microlensing is the study of the lensing due to the stellar mass objects in the lensing galaxy. Macroimages are comprised of a large number of microimages that are created by the stellar positions of individual stars in the lensing galaxy. The stellar positions, with separations on the order of micro-arcseconds, are impossible to determine, but can have a large impact on the magnifications of the macroimages. Macroimages from a lensing galaxy can be broken up into many microimages from pointlike stars at large optical depth \citep{Paczynski}. The effect of this microlensing at high optical depth can be very substantial, altering the observed intensity of the lensed images by 2 to 3 magnitudes \citep{Paczynski}. In fact, for quadruple lenses (gravitational lenses that create four lensed images of the source) microlensing is not only likely, but appears to be inevitable \citep{WittMaoSchechter}.

Anomalous flux ratios can often be accounted for by microlensing. A smaller ratio of stellar matter to dark matter actually increases, under some circumstances, the fluctuations in flux ratios caused by microlensing \citep{SchechterWambsganss}. Micro-saddle points may be both highly magnified or demagnified, while micro-minima may only be magnified to be greater than unity \citep{SchechterWambsganss}. 


\section{Monte-Carlo Microlensing Simulation}

\begin{figure}[htb]
\includegraphics[width=\linewidth]{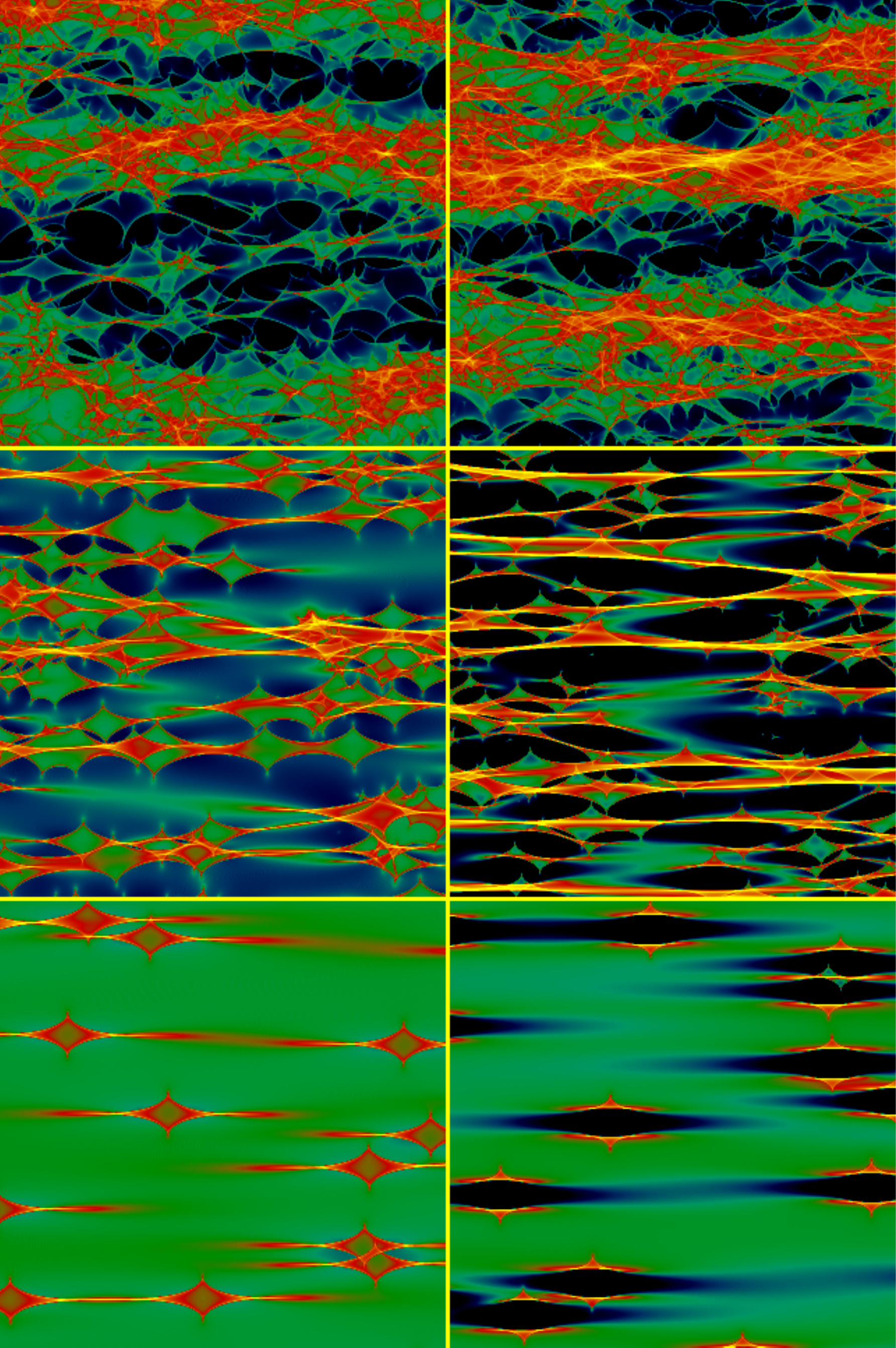}
\caption{A two-dimensional microlensing map for a typical macro-minimum (on the left) and macro-saddle (on the right) in the quasar plane. Uses stellar contribution to convergence ranging from $100\%$ in the top row, to $15\%$ in the middle row, and to $2\%$ in the bottom row. The total convergence is the same in all columns. The color scale ranges from dark blue (large demagnification from microlensing), to light blue, to green, to red, to yellow (large magnification from microlensing). Taken from \cite{SchechterWambsganss}.}
\label{fig:Wambsganss}
\end{figure}

\begin{figure}[htb]
\includegraphics[width=\linewidth]{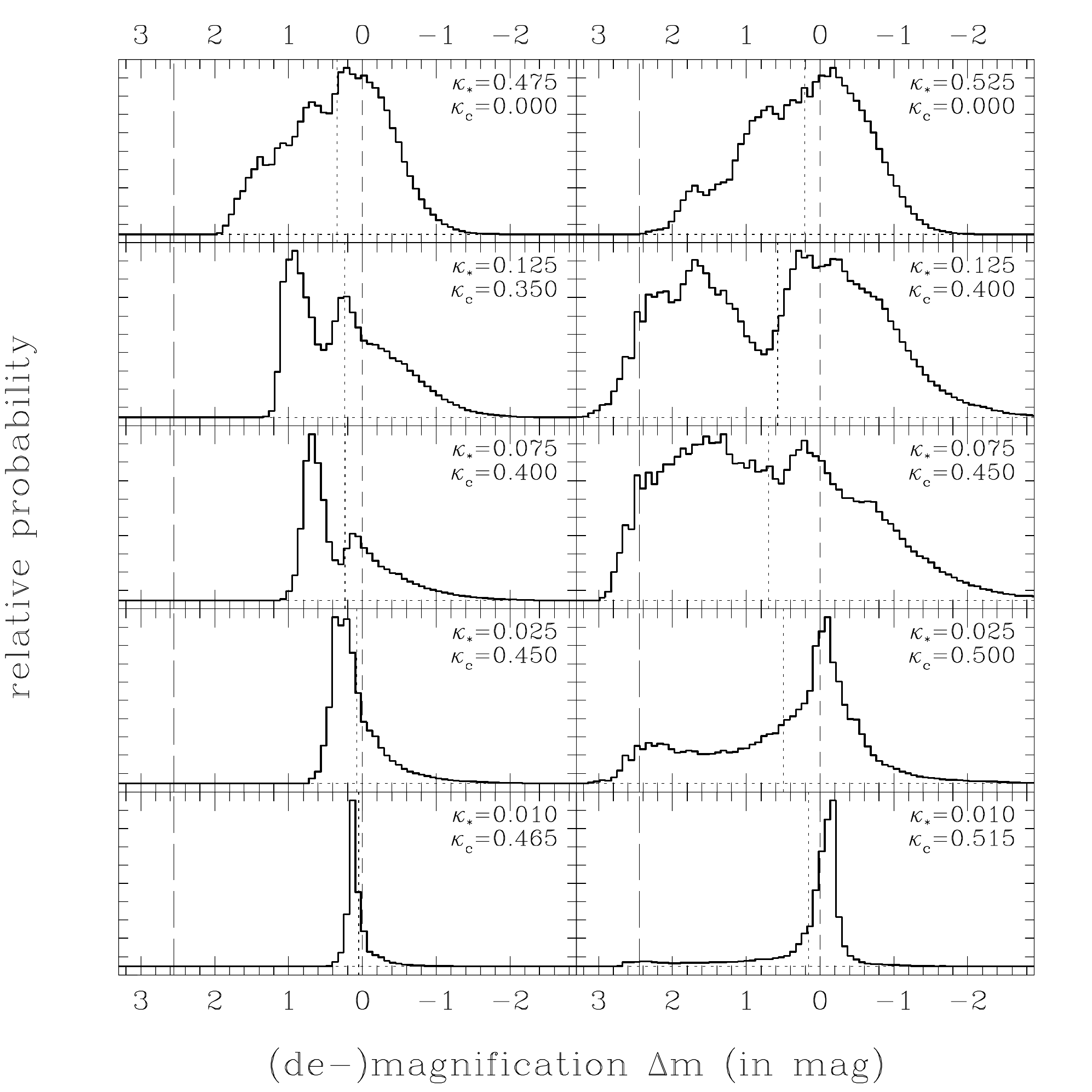}
\caption{Micro-magnification probability density for a macro-minimum (on the left) and macro-saddle (on the right). Uses stellar contribution to convergence ranging from $100\%$ in the top row, to $25\%$ in the second row, to $15\%$ in the middle row, to $5\%$ in the fourth row, and finally to $2\%$ in the bottom row. The total convergence is the same in all columns. The short-dashed line represents the theoretically expected macro-magnification, the dotted line represents the average magnification in magnitudes, and the long-dashed line represents the unlensed case. Taken from \cite{SchechterWambsganss}.}
\label{fig:Wambsganss2}
\end{figure}

\subsection{Microlensing Magnification Probability Density Maps}
Microlensing can be modeled using Wambsganss' inverse ray shooting code \citep{WambsganssPHD, Wambsganss}. Due to the mass sheet degeneracy \citep{Falco}, the three-dimensional space of shear ($\gamma$), convergence ($\kappa$), and stellar contribution to convergence ($\kappa^*$) can be projected on a two-dimensional convergence-shear ($\kappa ' - \gamma '$) space \citep{SchechterPooleyBlackburneWambsganss}. Doing so allows us to model the microlensing of a lensing system in a two-dimensional plane. The stellar contribution to convergence can also be changed in the model. The inverse ray simulation creates micromagnification probability density maps, which in turn can be used to create probability density curves for the four images of the lensed supernova \citep{Wambsganss}. Sample micromagnification maps, as created by inverse ray simulations, for a typical macro-minimum and macro-saddle point can be seen in Fig.~\ref{fig:Wambsganss}. The stellar contributions to convergence ranges from $100\%$ in the top row, to $15\%$ in the middle row, and finally to $2\%$ in the bottom row. The corresponding probability densities can be seen in Fig.~\ref{fig:Wambsganss2}, with the addition of rows displaying the probability densities for $25\%$ and $5\%$ stellar content.

\begin{figure*}[htb]
\includegraphics[width=\textwidth]{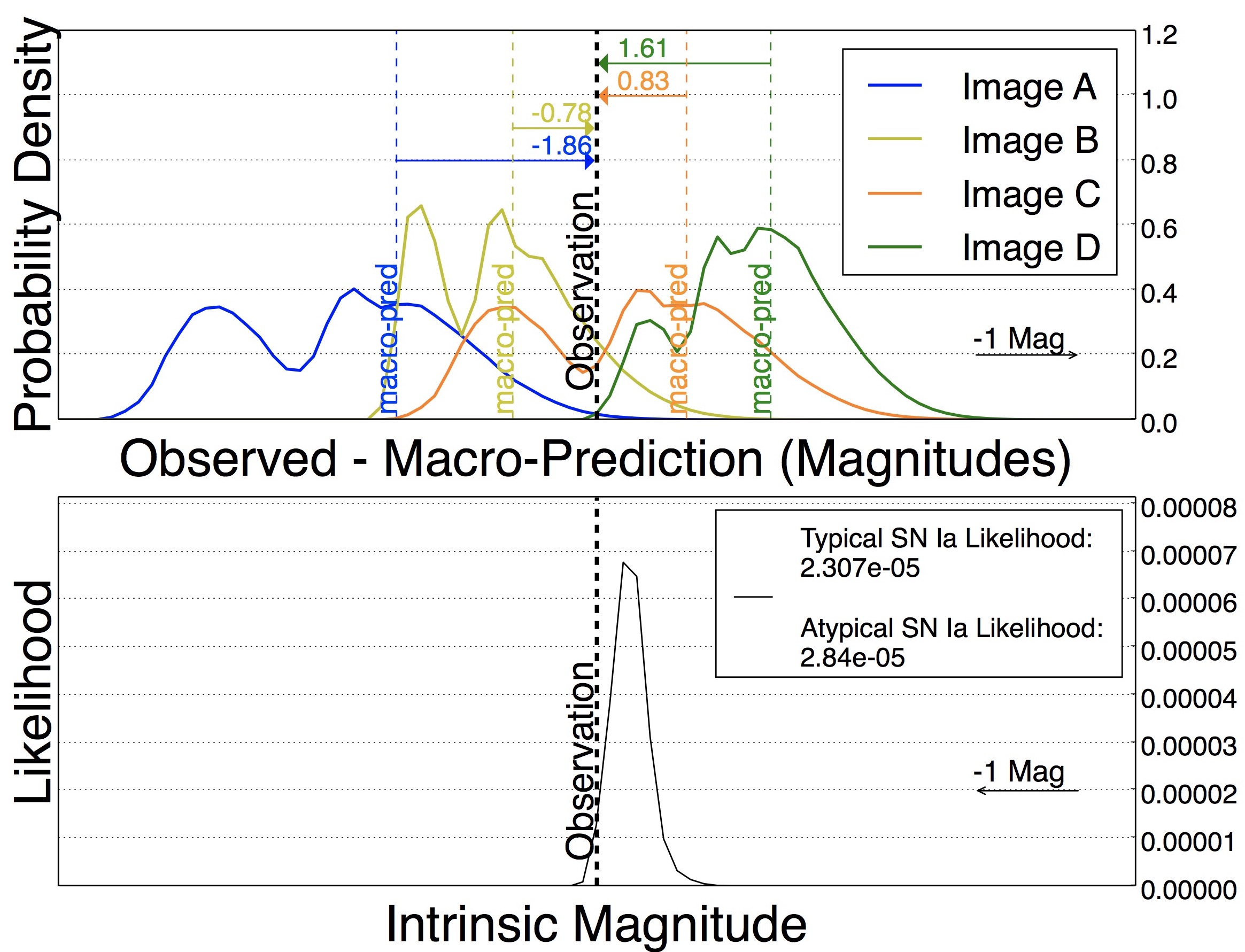}
\caption{Micro-magnification probability densities with $50\%$ 	stellar contribution to convergence for the four images of iPTF16geu in the top panel. The four probability densities have all been shifted such that the point of zero micro-magnification corresponds to the difference between the observed and macro-predicted flux values. By multiplying the four shifted probability densities, we create a plot that represents the likelihood for microlensing to cause the flux ratio anomalies between observations and macro-predictions as a function of the intrinsic magnitude of the source (bottom panel).}
\label{fig:MoreLikelihood}
\end{figure*}

As \cite{Pooley} demonstrated, analysis of the microlensing probability densities can be used to infer the likelihood for microlensing to cause the differences between the flux values that are predicted by macro-models and the flux values that are observed, at a given stellar to dark matter ratio. We generated the Wambsganss magnification maps for the four images of iPTF16geu using the convergence and shear values taken from \cite{More}. We then shifted the probability density curves for each image by the difference between its respective observed and predicted brightness magnitudes. This has the effect of setting the previous zero point for the four images as the ``observation,'' and the overlap of the four images probability densities represents the likelihood for micromagnification to cause the observed flux ratio anomalies. In analysis of the anomalous flux ratios of the supernova, we both assumed knowledge of the luminosity of the supernova and allowed the luminosity to be a free parameter. If we assume that we know the luminosity of the supernova, then in order to determine the likelihood for the microlensing to cause the observed flux ratio anomalies, we multiply the probability densities of the four images at the observation. We called this likelihood value the ``typical SN Ia likelihood.'' This represents the assumption that the supernova's brightness is well understood. However, if we cannot assume this, then, in order to make the luminosity a free parameter, we integrate over the total area under the overlap plot in order to obtain a value for the likelihood for microlensing to produce the flux ratio anomalies. This value was labeled the ``atypical SN Ia likelihood.'' The probability densities for the four images, and the resulting typical and atypical likelihood values, for $50\%$ stellar contribution to convergence, can be seen in Fig.~\ref{fig:MoreLikelihood}.

\subsection{Monte-Carlo Model}
A Monte-Carlo model can be followed in order to generate a distribution of standard microlensing likelihoods. In order to do so, four random numbers between one and zero are chosen. The four microlensing probability density curves for the four images are shifted such that the random number percentage of the area is to the left of the observation. By repeating this process ~1000 times, and keeping track of the microlensing likelihood for each of these sets of random microlensings, we can understand how likely is any one set of four microlensings for the system.

\begin{figure*}[htb]
\centering
\includegraphics[width=\textwidth]{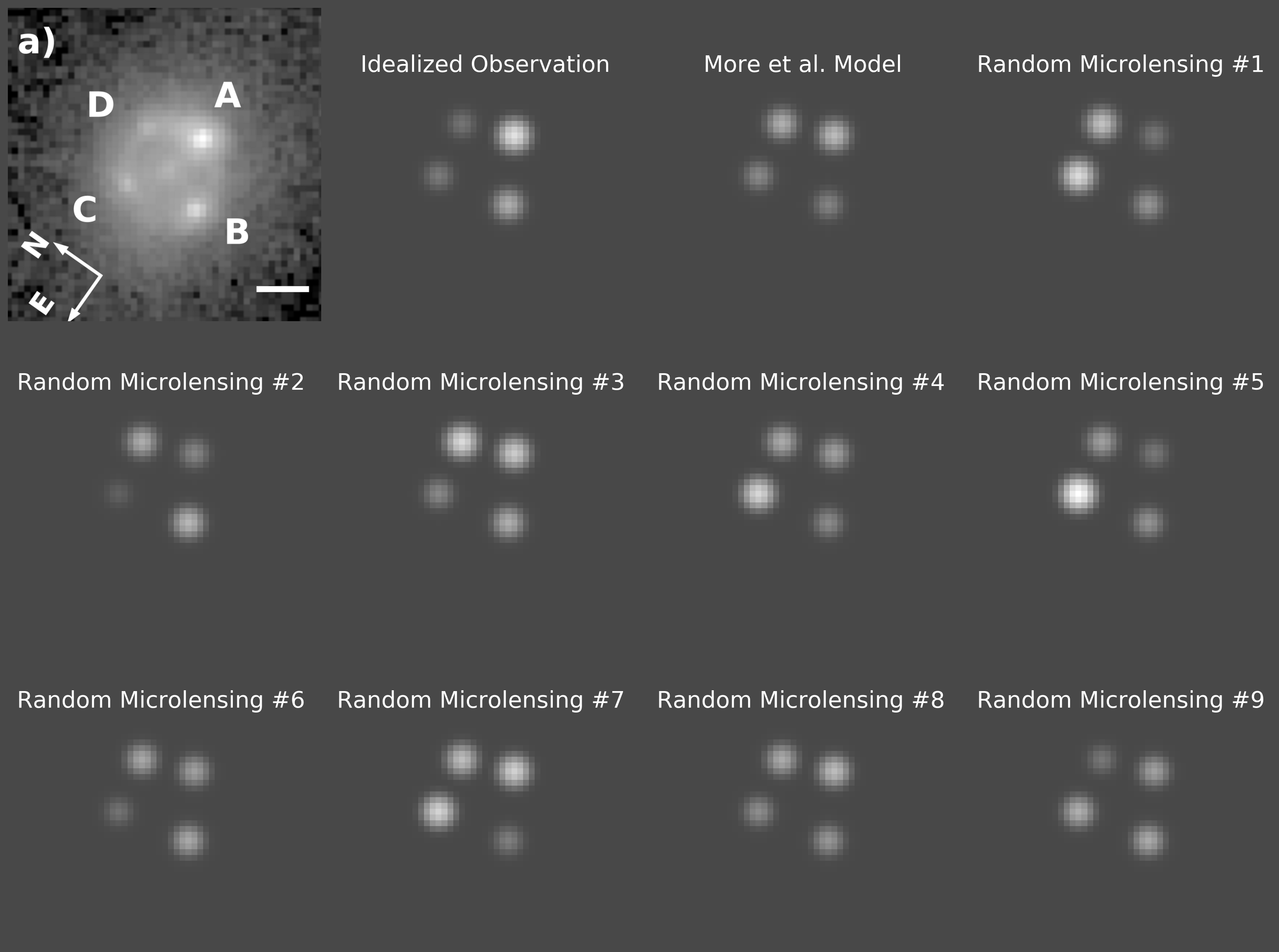}
\caption{Demonstrates the effects of microlensing on iPTF16geu. The top left image is taken from \cite{More}, and is an HST image (F814W) of iPTF16geu. The image titled "Idealized Observation" is our recreation of this image, taking each image to be Gaussian. The image titled "More et al. Model" is our recreation of the \cite{More} macro-predictions. Then, nine random microlensing values were determined for each lensed image, using the described Monte-Carlo model. The random microlensings were then added to the \cite{More} macro-model flux values. This figure demonstrates the significant impact of microlensing on the magnification of lensed images.}
\label{fig:ImageDemo}
\end{figure*}

Nine random sets of four microlensings were created using the Monte-Carlo process explained above The qualitative effects of these nine random microlensing are shown in Fig.~\ref{fig:ImageDemo}. In this figure, we were able to display the brightness predictions for a model containing both macrolensing and microlensing inputs for the images of the lensed supernova. As is evident in Fig.~\ref{fig:ImageDemo}, each of the nine random microlensings produces very different flux ratios, and none of the nine random microlensings produces flux ratios qualitatively similar to the observed flux ratios.

\section{Results}
\begin{figure*}[htb]
\includegraphics[width=\linewidth]{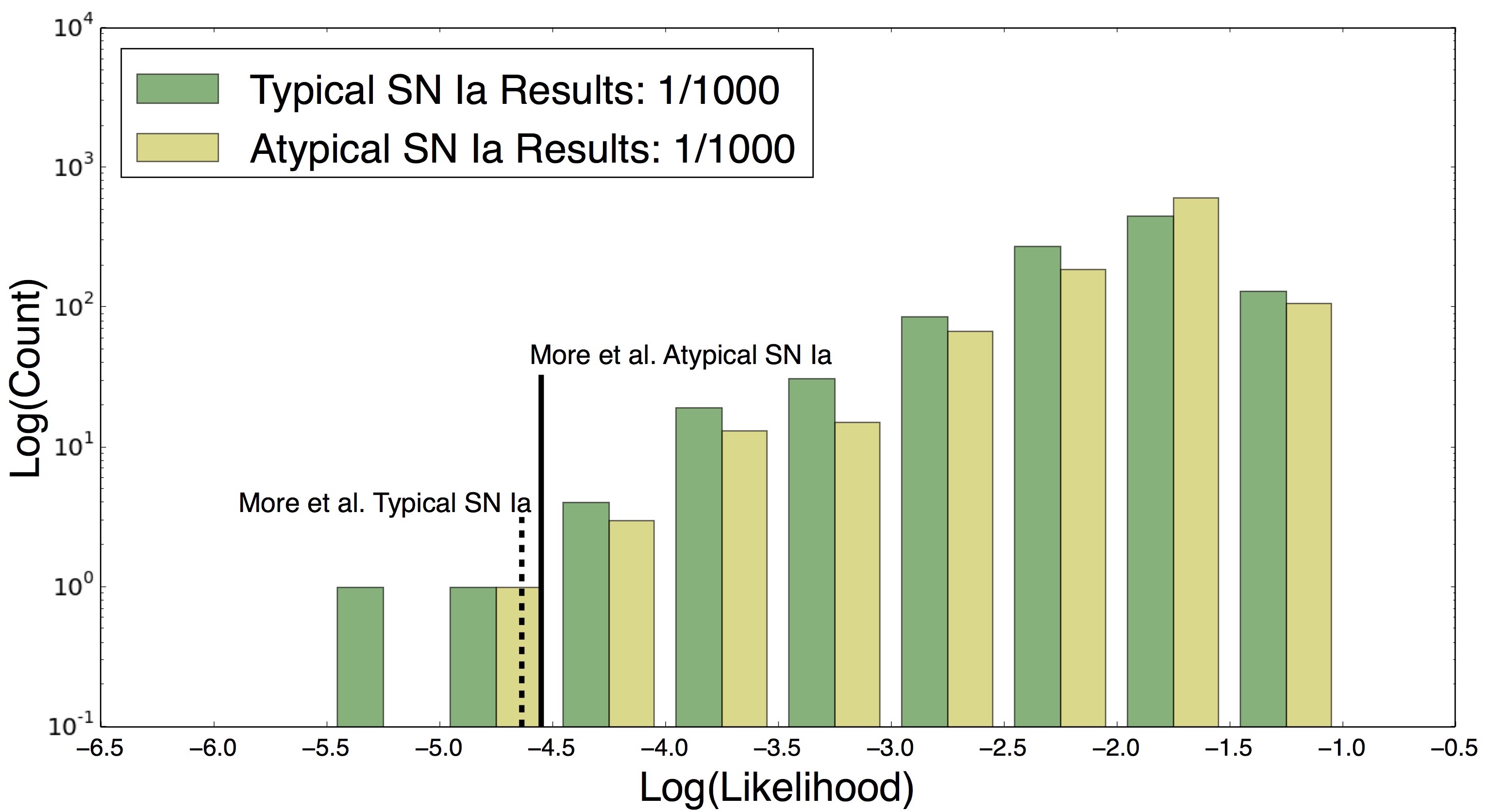}
\caption{Results of 1000 Monte-Carlo simulations at $50\%$ stellar contribution to the total mass. The black dashed line shows the typical SN Ia likelihood for the \cite{More} macro-model and the black solid line shows the atypical SN Ia likelihood for the \cite{More} macro-model. As only 1/1000 Monte-Carlo simulations produced a less probable typical and atypical microlensing likelihood than those required by the \cite{More} macro-models, it is evident that, at $50\%$ stellar contribution to the total mass, it is statistically very unlikely for microlensing to cause the flux ratio anomalies.}
\label{fig:results}
\end{figure*}

\subsection{Typical Likelihood}
We ran 1000 Monte-Carlo simulations at stellar contributions to total mass of $33\%$, $50\%$, and $100\%$, respectively. We found that at best the likelihood for microlensing to cause the flux ratio anomalies between macro-models and observations, if the object is a typical type Ia supernovae, is 3/1000. Fig.~\ref{fig:results} displays these results for both a typical and an atypical type Ia supernova at $50\%$ stellar contribution to the total mass.

\subsection{Atypical Likelihood}
We ran 1000 Monte-Carlo simulations at stellar contributions to total mass of $33\%$, $50\%$, and $100\%$, respectively. We found that at best the likelihood for microlensing to cause the difference between macro-models and observations, if the object is an atypical type Ia supernovae, is 2/1000. Fig.~\ref{fig:results} displays these results for both a typical and an atypical type Ia supernova at $50\%$ stellar contribution to the total mass.

\subsection{Different Macro-Models}
We then investigated alternative macro-models, specifically with varying contributions from external shear and ellipticity \citep{SchneiderKochanekWambsganss}. We ran 1000 Monte-Carlo simulations at stellar contributions to total mass of $33\%$, $50\%$, and $100\%$, respectively. We found that even with alternative macro-models, at best the likelihood for microlensing to cause the difference between macro-models and observations is 32/1000 (accounting for both the typical and an atypical likelihoods).

\subsection{Presence of Dust}
Further analysis of the data in \cite{Goobar} suggests that dust may play a significant role in the observation of image D. This is because there is a difference in image D's observed brightness at different wavelengths. Specifically, the HST F625W and HST F814W filters observed fluxes that were different by 0.74 magnitudes for image D. The presence of dust can have a large effect on the detected brightness of astronomical object, and could help to explain the flux ratio anomalies.

\subsection{What if the Macro-Models were Perfect?}
Even if the macro-models agreed perfectly with the observations, the FWHM of the intrinsic magnitude determined from the microlensing probability densities is 0.73 magnitudes. This stems from the wide spread of the microlensing probability densities, and results in a substantial difficulty in the ability to determine the intrinsic brightness of a lensed supernova with high accuracy. \cite{WambsganssSupernovae}, in analyzing the weak gravitational lensing on distant supernovae, also found significant lensing induced dispersions on truly standard candles.


\section{Conclusion}
We have examined the flux ratio anomalies between the macro-models and observations of the strongly lensed supernova iPTF16geu. We investigated the likelihood for microlensing to be the cause of the flux ratio anomalies, and found that it is statistically very unlikely to be the only cause. Through analyzing the full width half maximum, we determined that even if a perfect macro-model was found, it is difficult to determine the source object brightness with high precision, due to the effects of microlensing. The utility of lensed type Ia supernovae as standard candles is thus diminished by microlensing.




\printbibliography[title={References}] 


\end{document}